\documentclass[prl, twocolumn, footinbib,a4paper, showpacs, superscriptaddress]{revtex4}
\usepackage[latin1]{inputenc}
\usepackage{amssymb}
\usepackage{amsmath}
\usepackage{amsbsy}
\usepackage{bm}
\usepackage{graphicx}

\begin{document}

\title{ Two dimensional Leidenfrost Droplets in a Hele Shaw Cell }

\author{Franck Celestini}
   \affiliation{Universit\'e de Nice Sophia-Antipolis, CNRS, LPMC,
  UMR 7336,  Parc Valrose  06108 Nice Cedex 2, France}
   
\author{Thomas Frisch}
   \email{thomas.frisch@unice.fr}
  \affiliation{Universit\'e de Nice Sophia-Antipolis, CNRS,  INLN,   UMR 7335,   1361 Routes des lucioles, Sophia Antipolis F-06560 Valbonne France}

\author{ Alexandre Cohen}
    \affiliation{Universit\'e de Nice Sophia-Antipolis, CNRS, LPMC,
  UMR 7336, Parc Valrose  06108 Nice Cedex 2, France}
   
\author{ Christophe Raufaste}
 \affiliation{Universit\'e de Nice Sophia-Antipolis, CNRS, LPMC,
  UMR 7336,  Parc Valrose  06108 Nice Cedex 2, France}

    \author{Laurent Duchemin}
        \affiliation{ Institut de Recherche sur les ph\'enom\`enes hors-\'equilibre, CNRS UMR 7342, Aix-Marseille Universit\'e,  49 rue Joliot Curie, 13384 Marseille, France.}
        
  \author{Yves Pomeau}
        \affiliation{University of Arizona, Department of Mathematics, Tucson, AZ 85721 USA}

   
   

\begin{abstract}
We experimentally and theoretically investigate the behavior of Leidenfrost droplets inserted  in a Hele-Shaw cell. As a result of the confinement from the two surfaces, the droplet  has the shape of  a flattened disc and is thermally  isolated from the surface by the two  evaporating vapor layers. An analysis of the evaporation rate   using simple scaling arguments   is in agreement with the experimental results. Using the lubrication approximation we numerically determine the  shape of the droplets as a function of its radius.  We  furthermore find that the droplet width  tends to zero at its center when the radius reaches a critical value. 
This prediction is corroborated  experimentally  by the  direct observation   of the  sudden transition from a  flattened disc into an expending  torus.  Below this critical size, the droplets are also  displaying capillary azimuthal  oscillating modes   reminiscent of a  hydrodynamic instability.

\end{abstract}


\maketitle

\section{Introduction}

  In spite of its discovery in the late  19 th century \cite{leiden}, the Leidenfrost phenomenon is still today the subject of numerous studies for two  essential reasons. The first  reason is related to the strong decrease of the thermal  exchange  between the solid and the liquid due to the presence of  low thermal conductivity vapor layers:  this situation is of importance for example in metallurgy to control the cooling of metals \cite{bernardin} or in nuclear reactor safety \cite{vandam}. The second is  fundamental   and related to the fact  that a Leidenfrost droplet  may be considered as  an ideal realization of a perfect non-wetting system \cite{clanet,quere}. These droplets have shown  rich and unexpected behaviors \cite{quere}. For example  drops on periodic patterned surface display a drift due to the spatial  symmetry breaking. Furthermore possible applications of Leidenfrost droplets might be the transport of liquid in the milli-fluidic  or micro-fluidic area \cite{tabeling}.  For example low pressure  Leidenfrost droplets have been shown recently  to be stable at room temperature and could be  subsequently potential receptor of particles in solution \cite{pression}.
Surprisingly, to our knowledge no studies have been devoted yet to the Leidenfrost effect in a  2d confined  geometry. This is somehow  unexpected because such a complex phenomenon should display  new properties as  the spatial dimension is reduced.  The aim of this paper is  thus to investigated a simple situation representative of the effect of  spatial confinement on the Leidenfrost droplets. These droplets are  inserted in a horizontal  Hele-Shaw cell whose gap is smaller than the capillary length $\kappa^{-1} = \sqrt{\gamma/\rho_l g}$ with $\gamma$ being  the surface tension, $\rho_l$  the liquid density and  $g$ the acceleration of gravity. As a result of the confinement from the two surfaces the drop  takes the shape of  a flattened  saucer-like disc which floats  between two vapor layers. These drops are quasi thermally 
 isolated from the surface by the  evaporating vapor layers and they display  undulating star-like shapes. In this Article, we first describe our experimental set-up  and we  describe the dynamic of evaporation by a simple model  which takes into account the Poiseuille flow  in the vapor layer. This model is in agreement with the experimental results for the global evaporation rates.  We then discuss the vertical profile of the drop using a theoretical model  which results from the balance  between   surface tension  and the Poiseuille  flow in the vapor layer. We  observed  that the droplets have a maximum radius $R_c$ beyond which they transform into a torus by the process of a  hole nucleation  and expansion at their center. These  experimental results  are confronted with our  theoretical model and fall within a  good agreement. We also report observations of  large amplitude star-like undulation  consisting of  azimuthal  oscillating capillary wave. The frequency of  oscillations is measured and is found to be close to the frequency of  Rayleigh  capillary wave  of droplets.
Finally, we  discuss   possible   mechanisms at the origin of the instability leading to star-like oscillations of the droplets.

\begin{figure}
 \includegraphics[width=.4\textwidth]{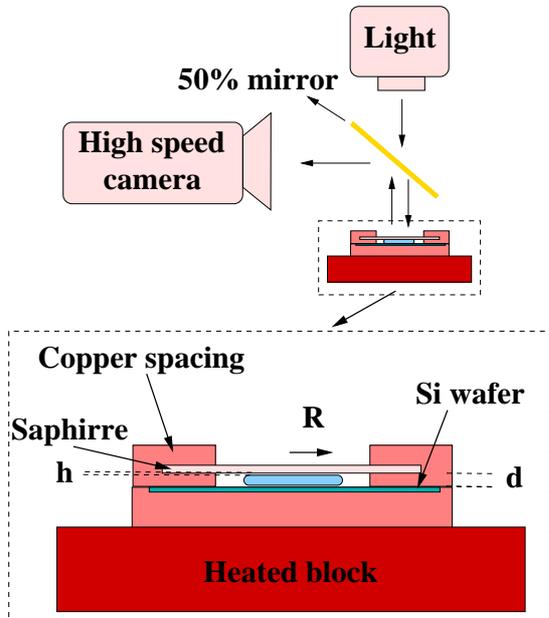}
 \caption{Experimental set-up : a Hele-Shaw cell is heated at a controlled temperature $T_P$. A Leidenfrost droplet of radius $R$ is inserted between the two hot plates separated by a spacing $d$.}
 \label{setup}
\end{figure}

\section{Evaporation dynamic }

\subsection{Experimental results}

Our experimental set-up is depicted in Fig.\ref{setup}. A heated copper block permits to control the temperature $T_p$ of two plates separated by a spacing of width $d$. This temperature is measured with a PT100  temperature sensor. The upper plate is made of sapphire whose optical and thermal properties  permit us to visualize the droplet from the top through a semi-transparency mirror. The lower plate is covered with a silicon wafer. Images are recorded with a high speed camera with a  frame rate  varying between  $60$ and $4000$ frames per second.
As for classical Leidenfrost droplets we suppose that the liquid is at its boiling temperature $T_b$ and we denote $\Delta T = T_p-T_b$ the temperature difference between  the hot plates  and the droplet. For all the experiments, the plate temperature has been fixed to $T_p=300 $ Celsius, giving a value of $\Delta T = 200$ Celsius. A capillary is used to insert the ultra distilled water droplet  into the Hele-Shaw cell. Different spacings  $d$ have been used ranging between $d=0.3$ and $2$ mm.

\begin{figure}
\includegraphics[width=.4\textwidth]{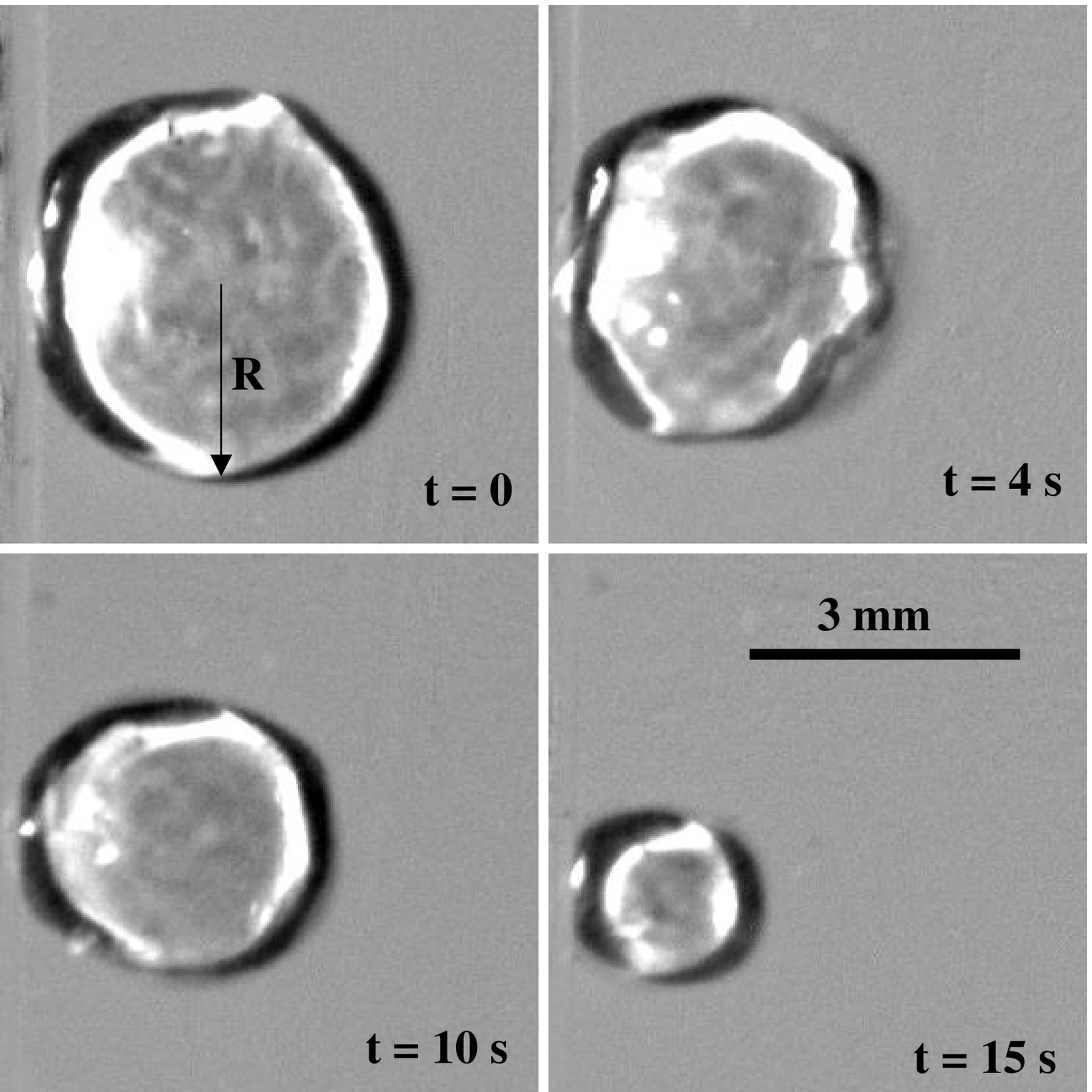}
 \caption{Snapshots of an evaporating droplet (top view) at different time. The spacing of the Hele-Shaw cell is of $1$ mm and the temperature of the plates is $T_p = 300$ Celsius. }
 \label{snapshot}
\end{figure}

We first investigate the dynamics of the evaporating droplets.  The lifetime of the droplet is about  ten seconds, one order of magnitude lower than usual Leidenfrost droplets and two orders of magnitude lower  than low pressure Leidenfrost droplets \cite{pression}. As illustrated on Fig. \ref{snapshot}, the droplet  is  slowly evaporating and its radius  $R$ is decreasing with time. An image analysis permits to record the evolution of $R$ with the time $t$,   its value is deducted from the apparent area of the droplet viewed from the top. It is worth noticing that the shape of the drop is not necessary circular but presents some contour oscillations that will be discussed in the last section of this Article.
We represent in Fig. \ref{rdt} the  time evolution of  the radius $R$ for droplets inserted in three cells with different spacings : $d=0.3$, $0.5$ and $1 {\rm  \, mm}$. We clearly see that the lower the spacing the faster the evaporation of the droplet. The full lines correspond to a best-fit   of  the phenomenological model  which is presented just below.

\subsection{Phenomenological model}
\begin{figure}
 \includegraphics[width=.5\textwidth]{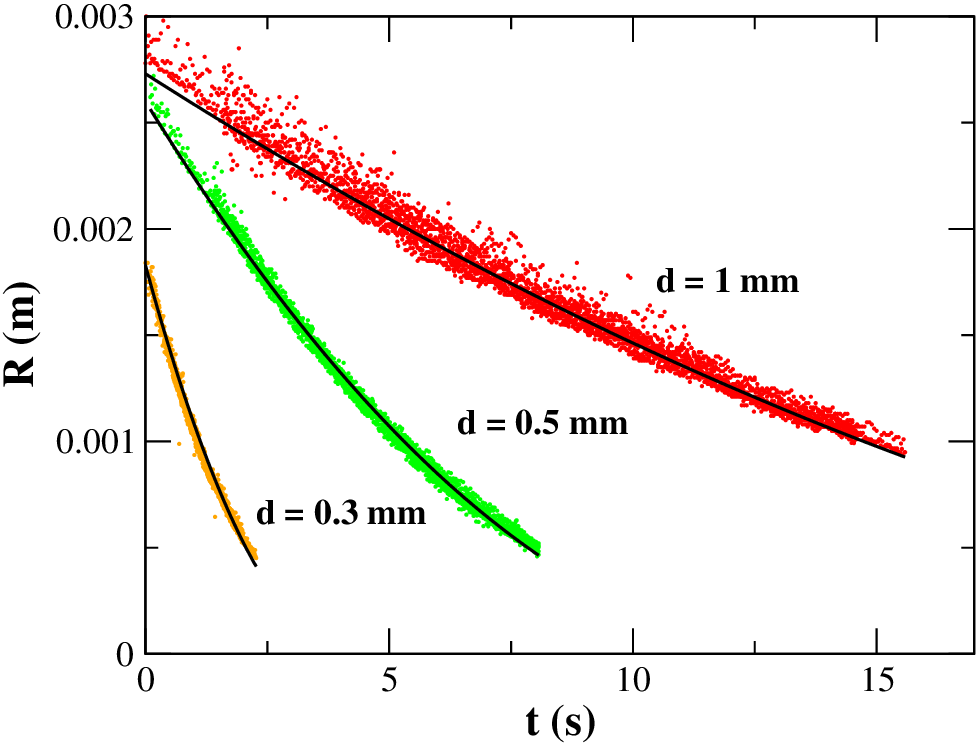}
 \caption{Droplet radius as a function of time for three different spacings $d=1$,  $0.5$  and  $d=0.3$ mm respectively in red, green and brown circles. The full lines correspond to the best fit to Eq. (\ref{rdteq}) }
 \label{rdt}
\end{figure}
One of the difficulties in the Leidenfrost system consists in the determination of the thickness $h$  of  the vapor  film which is situated between the drop and the heated plates. 
In order to estimate $h$ we develop a model  based on a  simple scaling argument which applies to a flat droplet of radius $R$ as shown on Fig. 1. 
The vapor film thickness results  from the balance between the  pressure  (driven by the  Poiseuille flow  produced by the local vapor evaporation)  and the capillary surface tension  effects.
 Energy conservation during the evaporation  process  (namely Stefan's boundary condition on the liquid-vapor interface) and Fourier law  for the heat transfer in the film yield the order of magnitude for the vertical velocity  $w$  of the vapor near the surface of the droplet. It  reads:

 \begin{equation}
 \rho_v L w  \propto \lambda \Delta T /h \, .
\label{eq1}
 \end{equation}
Here $ L$ is the  latent heat per unit mass,   $\lambda$ is the thermal diffusivity coefficient, $\Delta T$ is the temperature difference between the  hot plates  and the droplet and $\rho_v$ is the density of the vapor.
Mass conservation
 yields the following estimate for the magnitude of the horizontal velocity $u$ in terms of the vertical  velocity $w$:
\begin{equation}
u /R \propto w/h  \, .
\label{eq2}
 \end{equation}

Assuming a Poiseuille flow in the vapor  in the  lubrication regime leads to an  horizontal  pressure drop   $\Delta P$    which scales as
\begin{equation}
 \Delta P /R \propto \eta u / h^2  \, ,
 \label{eq3}
  \end{equation}
where $\eta$ is the  vapor film viscosity.

 Finally,  we note that the Laplace pressure  $\Delta P$ between the drop and the gas  reads:
 \begin{equation}
 \Delta P   \propto \gamma/ d \, ,
 \label{eq4}
 \end{equation}  Here we have neglected the small curvature term which  is smaller by a factor $d/R$.
 
  We obtain using Eqs. (\ref{eq1}-\ref{eq4})  a relation for the film height $h$ which reads:
 \begin{equation}
 h    \propto R^{1/2} d^{1/4} \left(\frac{\eta \lambda \Delta T}{\rho_v L \gamma }\right)^{1/4} \, ,
 \label{eq5}
 \end{equation} which can be rewritten simply as :
 \begin{equation}
 h   \propto R^{1/2} d^{1/4} l^{1/4} \, ,
 \label{eq6}
 \end{equation}
 where we have introduced a  characteristic  length $l$  defined  as :
   \begin{equation}
  l=\frac{\eta \lambda \Delta T}{\rho_v L \gamma } \, .
  \end{equation}
As shown by Eq. (\ref{eq6}),  the width of the vapor layer  $h$ can thus be deduced from  the measured value  $R$.  
  
 Furthermore we can   also verify experimentally that the previous scaling holds by measuring the time evolution of the radius of the drop $R(t)$. The rate of evaporation of the droplet can be deduced from the outgoing flux of vapor and reads:
  \begin{equation}
 \rho_l \frac{d}{dt}( \pi R^2 d)   \propto 2  \rho_v \pi R^2 w  \, .
 \end{equation}

Solving this differential  equations using  Eqs. (\ref{eq1}) and (\ref{eq6}) we obtain  the evolution  radius $R(t)$ as:
 \begin{equation}
 R(t)   \propto  (R_0^{1/2} - C t)^2 \, .
 \label{rdteq}
 \end{equation}
Here
\begin{equation}
 C = \frac{\lambda \Delta T}{\rho_l L l^{1/4}} d^{-5/4} \, .
 \label{ceq}
 \end{equation}

\begin{figure}[!h]
 \includegraphics[width=.5\textwidth]{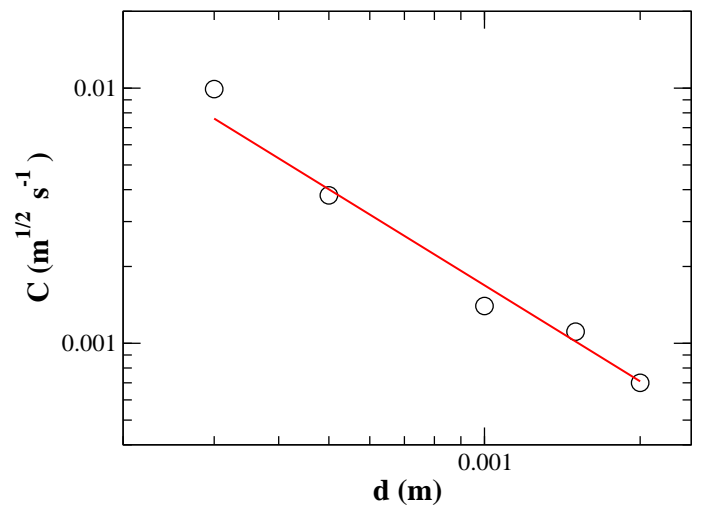}
 \caption{Values of the parameter $C$ as a function of the spacing $d$. The full line corresponds to the scaling law predicted in Eq. (\ref{ceq}). }
 \label{c}
\end{figure}

 As shown on Fig. \ref{rdt},  we compare   the  experimental measurements of the  time evolution  of the radius $R(t)$   with the theoretical prediction  given in  Eq. (\ref{rdteq})  for three different values of the spacing $d$.
  A satisfactory agreement is found between  the theory and the experiments. We can infer the value of  the constant  $C$  from a best-fit adjustment of the data given on Fig. \ref{rdt} using the  time evolution for $R(t)$ given in   Eq. (\ref{rdteq}).  As shown on Fig. \ref{c} the values of $C$ obtained lead to the  expected  dependence $C\propto
d^{-5/4}$ as predicted by  Eq. (\ref{ceq}). It is worth noticing that the  value of the pre-factor $ \frac{\lambda \Delta T}{\rho_l L l^{1/4}}$  found experimentally  using the data shown on Fig. \ref{rdt}   is  $2.9 \times 10^{-7}  {\rm m} ^{1/2} \,  {\rm  s}^{-1} $. It  is of  same order than the one predicted  $  5.4 \times 10^{-7}  {\rm  m} ^{1/2} \,  {\rm  s}^{-1} $.  \\Here we have used the following values for the physical parameters:
$\lambda = 0.032   \,  {\rm J} \,  {\rm s}^{-1}  {\rm K}^{-1}$, $L = 2.25 \,  10^6 {\rm J}  \, {\rm kg}^{-1}$, $\rho_v = 0.81 {\rm kg} \, {\rm m}^{-3}$, $\eta = 1.3  \, 10^{-5} {\rm kg} \, {\rm m}^{-1} \,  {\rm s}^{-1}$, $\kappa = \lambda/(L \rho_v)$, $\Delta T = 200  \,{\rm  Celsius} $, $\sigma = 0.072 \,   {\rm J} {\rm m}^{-2}$ and $d = 0.001{\rm m}$

\section{Droplet  shape and hole nucleation}

\begin{figure}[!h]
 \includegraphics[width=.5\textwidth]{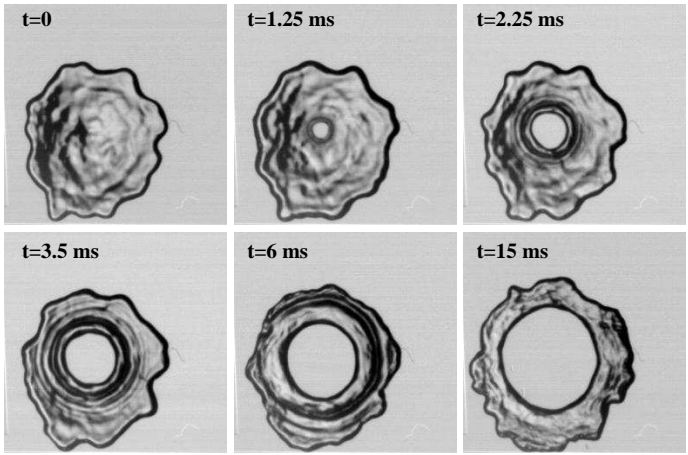}
 \caption{ Images sequence of the hole nucleation and growth.}
 
 \label{explo}
\end{figure}
The previous analysis is based upon a strong hypothesis of a constant vapor thickness, both below and above the droplet. It has been recently shown that this hypothesis does not hold properly for 3D Leidenfrost droplet \cite{CRAS}. Therefore we decided to perform  a deeper  theoretical analysis to predict the vapor layer profile in our confined geometry.

\subsection{Lubrication model}

\begin{figure}[!h]
 \includegraphics[width=.5\textwidth]{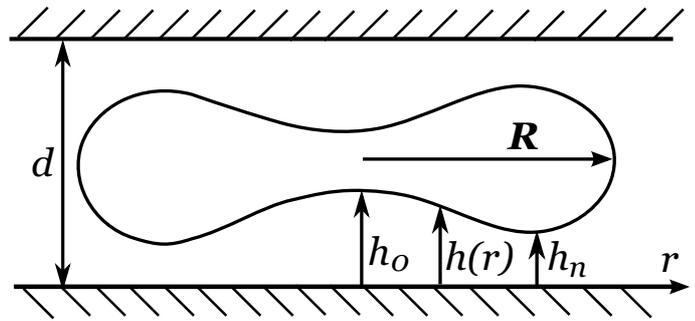}
 \caption{sketch of a vertical cut of a drop. $h_0$, $h_n$  and $R$ are shown. }
 \label{sketchvertical}
\end{figure}

Let    the  height of the vapor layer at the equator be   $h_0$, the neck height $h_n$ and the radius of the droplet be $R$ as shown on  Fig.  \ref{sketchvertical}.
The two plates   are held at  temperature $T_p$  and are separated by a distance $d$,  the  temperature of the drop is $T_b$.
Let $h(r)$  be the shape  on the lower interface and $d-h(r)$  the shape of the upper interface.The balance  between the surface tension effect and the Poiseuille flow  leads to a variety of shapes  which can  described  using the lubrication approximation. This approach has  been used successfully in another context such as lens floating \cite{duchemin}.
The pressure  $p$ in the vapor film just outside the drop is driven by the following equation as shown in Appendix A  and reference  \cite{CRAS}:
\begin{equation}
 \partial_r \left(\frac{ r h^3 }{12} \partial_r p\right)+\frac{ r \eta \kappa \Delta T}{h(r)}=0 \, ,
 \label{eqpomeau}
\end{equation}
where   $\kappa= \lambda /(\rho_v L)$. Equation (\ref{eqpomeau}) derives from the incompressibility condition of the  vapor and the Poisseuille hydrodynamic  relation  \cite{CRAS}.
The pressure  in the vapor film just below the drop is given by:
\begin{equation}
p=p_d-\sigma \kappa \, .
\label{eqpression}
\end{equation}
Here $p_d$ is the pressure in the drop which is supposed to be constant,  $\sigma$ is the surface tension and $\kappa$ is the  mean curvature which reads:
\begin{equation}
\kappa= \partial_s \psi+\sin{\psi}/r \, .
\label{eqcourbure}
\end{equation}
Let us choose  $d$ as the  unit of length and $\sigma/d$ the pressure unit. in these units  Eq. (\ref{eqpomeau})  reads:
  \begin{equation}
 \partial_r \left(\frac{ r h^3 }{12} \partial_r p\right)+\frac{ r \alpha}{h}=0 \, ,
 \label{equadim}
\end{equation}
where  $\alpha= \frac{\eta \kappa \delta T}{\sigma d}$    is a dimensionless parameter which depends on $d$.

 The lubrication  Eq.  (\ref{eqcourbure}, \ref{equadim})  can  be  reformulated   in curvilinear coordinates as:
 \begin{eqnarray}
\partial_s h & = & \sin{\psi}   \label{eqshootdep}  \, ,\\
\partial_s r &  = & \cos{\psi}  \, ,\\
 \partial_s \psi & = & \kappa- \sin{\psi}/r \, ,\\
 \partial_s \kappa & = & \kappa_1 \, ,\\
 \partial_s \kappa_1 &= &\frac{12 \alpha  \cos ^2(\psi)}{h^4}-\frac{3 \kappa_1h'}{h}  \nonumber \\
 & & - \frac{  \kappa_1 \left( r  \psi ' \tan (\psi)+r' \right)}{r}   \, .
 \label{eqshoot}
 \end{eqnarray}

Here we have used the following geometric  transformation $\partial_r =\partial_s  \frac{\partial s}{\partial r}=\frac{\partial_s}{ \cos(\psi)} $.  The boundary conditions at $r=0$ read:
\begin{eqnarray}
r(s=0) &=& 0 \, , \\
h(s=0) & =& h_0 \, , \\
\psi(0)  & = &  0 \, , \\
\kappa(0)& =&  \kappa_0 \,  ,\\
\partial_s \kappa(0)&=& 0 \, .
\end{eqnarray}

The boundary conditions at $s=s_{max}$ which corresponds to the equator  read:
\begin{eqnarray}
r(s=s_{max}) &=& R\, , \\
h(s_{max}) & = & 1/2 \\ 
\psi(s_{max})  & = &  \pi/2\\
\partial_s \kappa(s_{max})&=& 0
\end{eqnarray}

\begin{figure}[h]
\includegraphics[width=.4\textwidth]{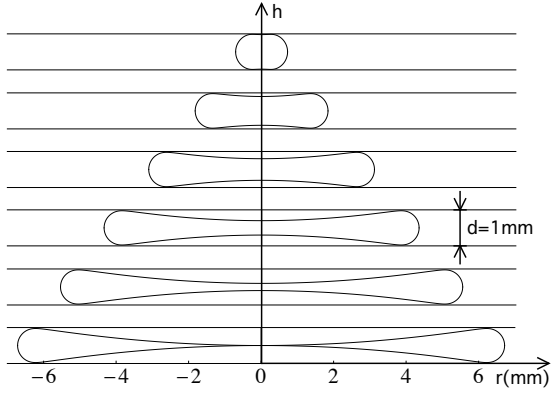}
\caption{Profile of a drop obtained by  numerical resolutions of Eqs. (15-19)  for  6 different values o the parameters  $h_0$.  Top $h_0=0.001 {\rm mm}$, bottom $h_0=0.5  {\rm mm}$.
 The horizontal axis is $r$, the vertical axis  is the  $z$ axis. The vertical  distance  $d$ between the two plates is  $1$ mm. The bottom curve represent the the curve for which $R=R_c$. }
\label{familly}
\end{figure}

\begin{figure}[!h]
\includegraphics[width=.5\textwidth]{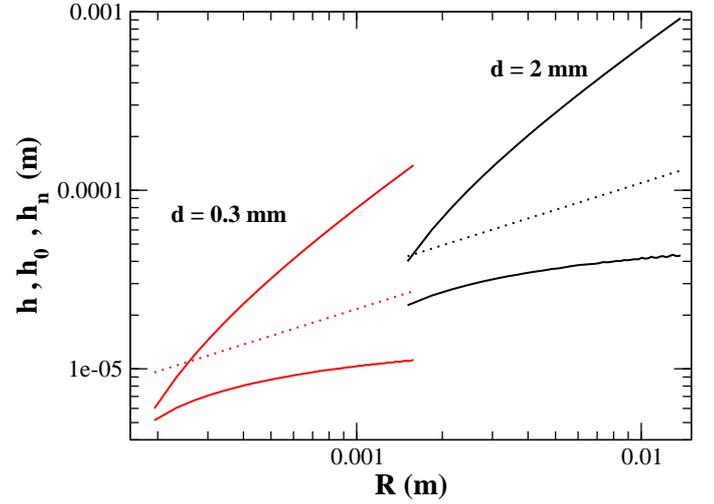}
\caption{Numerical simulation of Eqs. (15-19). We plot  $h_0$  and $h_n$  (full lines respectively top and bottom)   as a function of $R$.  The red curve is   for  $d=0.3$ mm,  the black  is for  $d=2$mm.  The  dashed-lines represent the scaling $ h  \propto R^{1/2}$ predicted by Eq. (\ref{eq5}) using the data deduced  from  the experimental measurement of the constant $C$. The prefactor   in Eq. \ref{ceq} are estimated using a best-fit adjustment of the results show on Fig. 3.  }
 \label{figsimu}
 \end{figure}

There are two free parameters
$ \kappa_{o}$ and $s_{max}$ which need to be adjusted in order to satisfy the boundary conditions defined in Eqs. (27) and (28).
These equations can be solved using a  standard shooting method. As shown in Fig. \ref{familly} we represent   the shape of the drop for different values of the $h_{0}$,
we find that there exists a critical radius $R_c$ beyond which  a simply connected drop cannot exist as shown for example on the  bottom profile displayed  on Fig. \ref{familly}.
Even though the direct measurement of vertical profile of the droplet are not possible, we have estimated the value of the vapor layer $h$ by proceeding  in the following manner.
Using the relation obtained in Eq. \ref{eq5} we can estimate the value of  $h$ by a  measurement of $R$. Here the  value for the pre-factor in Eq. \ref{eq5}
 is deducted from  the experimental measurement of the constant $C$,  the pre-factor   in Eq. \ref{ceq} is estimated using a best-fit adjustment of the results shown on Fig. 3.
 Finally we display on  Fig. \ref{figsimu}    the value of  $h_0$ and $h_n$  obtained by numerical simulation of Eq. (15-19) as a function of $R$ for two values of the  spacing $d$. We superpose  on these  curves 
the value of  the thickness  $h$  estimated above. As expected the values of $h$ ranges between $h_{neck}$ and $h_{0}$.


\subsection{Hole nucleation}

\begin{figure}[!h]
\vspace{1.5cm}
 \includegraphics[width=.5\textwidth]{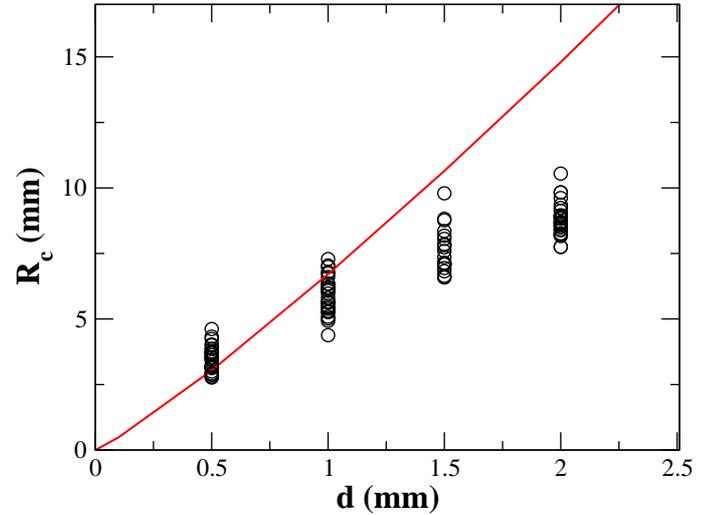}
 \caption{Critical radius $R_c$ versus the spacing $d$. Each point corresponds to  one experiment. The red line holds for the numerical prediction using Eqs. (\ref{eqshootdep}-\ref{eqshoot})}  .
 \label{rc}
\end{figure}

The numerical resolution of the droplet profile as shown in the above section  reveals  that no solution should exist   beyond  a critical drop radius $R_c$. This statement is thus tested experimentally: we first inject a primary drop in our system that is grown continuously by coalescence with smaller drops. We find that at some point a hole nucleates and grows inside the drops (Fig. \ref{explo}).  The hole grows until it reaches the drop edges.  This leads to the destruction of the drop into several fragments that are ejected radially. The critical radius $R_c$ exhibits some dispersion, but a clear trend can be observed as a function of the spacing $d$ (Fig. \ref{rc}). The comparison with the model is rather satisfactory given that there is no free parameter. The agreement is even very good for smaller spacing. For larger spacing we might expect  some discrepancies since  gravitational effects could play a role as the spacing becomes comparable to the capillary length $l_c=\sqrt{\sigma/\rho g}$

\section{ Oscillation modes }
In our experiment, the drops were found  to have  radial oscillations mode for  most of the values of the  parameters  such as the  radius $R$  and  the spacing between the  two plates $d$.  Theses oscillations can be  described by star-shaped contour modes with   frequencies ranging between 10 and 400 Hz. The modes are thus parametrized by a integer $m$, namely the azimuthal  mode number, which measures the number of  spikes along the  contour of the drop as shown on Fig.  \ref{mode1}. As shown for example on Fig. {\ref{newfig}, we display the evolution of the droplet for the   $m=3$ mode   during  a half-period, this oscillation is typical of a standing wave pattern.

\begin{figure}[!h]
 \includegraphics[width=.4\textwidth]{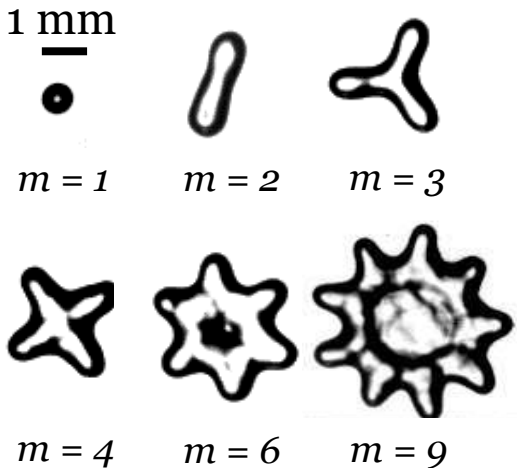}
 \caption{Top views of the droplets. Examples of   different oscillation modes.  Here  $m$ holds for the number of spikes of the drop, $d=  1 {\rm mm}$ and the horizontal scale is shown by the black bar.}
 \label{mode1}
\end{figure}

\begin{figure}[!h]
 \includegraphics[width=.4\textwidth]{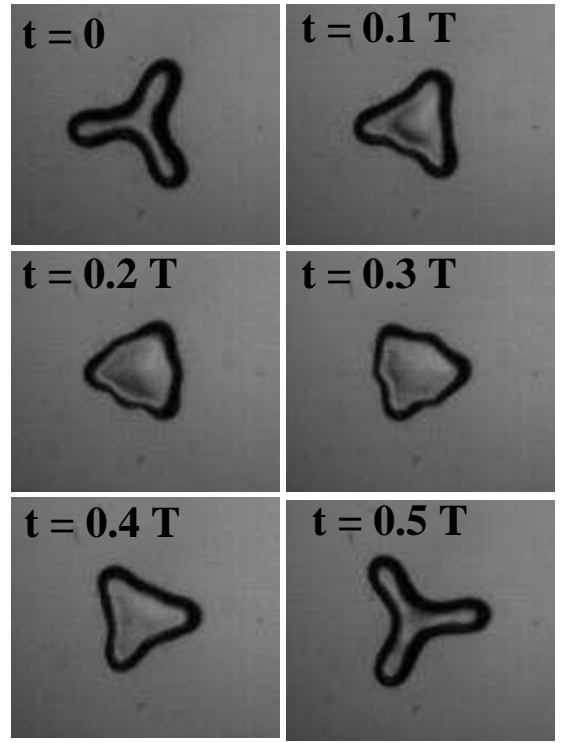}
 \caption{Top view of a single drolet. Time evolution of the  $m=3$  mode for a half-period. Here $d=  1$ mm, $R=1$ mm   and ${\bf{T}}=0.009 \,  s$.}
 \label{newfig}
\end{figure}

During a typical experiment in which drops slowly evaporate,  we observe that the azimuthal number  $m$   has  tendency to decrease with time   as the drop size  decreases.
However, there seems to be no stringent dynamics for  the  time evolution of the azimuthal  mode number and in particular a wide range of frequency transitions occurs during the life of a  droplet.
Furthermore  there  seems to be no  direct link between the azimuthal  number $m$ and the droplet radius except for  this  decreasing  tendency. The value of  the mode number  displays  strong  stochasticity probably induced by the non-linear coupling between oscillations modes and  by the inherent thermal fluctuations  which are  expected at  liquid-vapor interfaces just below the boiling transition.\\
We plot on   Fig. \ref{mode2}  the frequency  measurements with respect to the radius $R$   for different values of $m$. They reveal a clear trend well approximated by  a power law as explained below.
Let us  first focus for example  on  the mode  $m= 3$ which was observed over the larger range  of frequencies and for several  values of the spacing $d$. 
We did not notice any effect of the spacing so that we performed a power law interpolation over all its data. This leads to $f$(Hz)$= (12\pm 1) \times 10^{-4} R$(mm)$^{-1.7 \pm 0.3}$. Even though other modes $m$ do not span as much  frequencies as  the mode $m= 3$ does,  their trends are also compatible with a scaling of $R^ {-3/2}$. Such an exponent is therefore reminiscent of the Rayleigh spectrum for capillary wave on spherical drops  \cite{Lamb1932}  or two dimensional discs  \cite{Takaki1985} for which the  frequency also scales likes $R^{-3/2}$ . In order to go deeper in this analysis we have plotted on Fig. \ref{mode3}  the dimensionless frequency  $f/f_0$ of the contour modes as a function of $m$,
 where $f_0=\frac{1}{2\pi} \sqrt{\frac{\gamma}{\rho R^3}}  \, .$ is the typical frequency scale. The  comparisons between the experimental  points and the theoretical results  shown on Figure  \ref{mode3}  seem to validate the capillary origin of these surface waves.   The  small discrepancy  emphasized by the slight overestimate of the theory may  origin from several facts such as  the complexity of the Leidenfrost effect which involves vapor flow  around the droplet, the  possible coupling between the gas dynamics and the fluids at the surface and the particular  geometry of a droplet in a Hele-Shaw cell.\\ 

\begin{figure}[!h]
 \includegraphics[width=.5\textwidth]{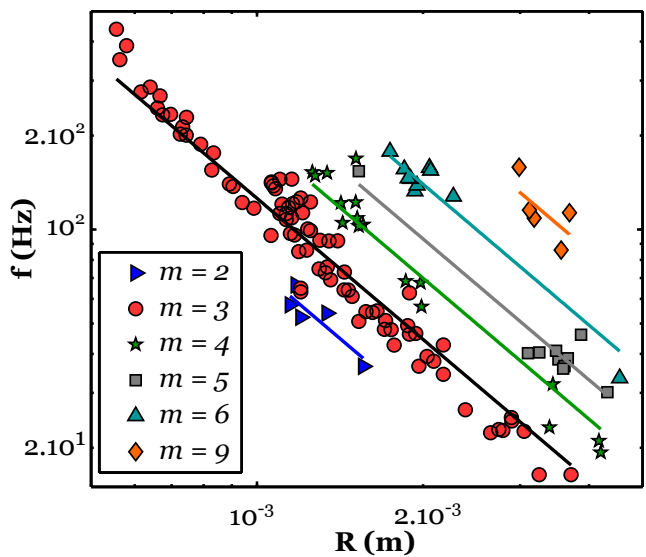}
 \caption{
Oscillation frequency $f$ versus the drop radius $R$ in meters. Several modes are represented. Each of them is fitted by a power law with an exponent -3/2 (solid lines). Experiments are performed for $d = 1 $mm, excepted  for $m=3$, where data of $d=0.5$ and 0.3 mm are also plotted.}
 \label{mode2}
\end{figure}

The comparison of our experimental results  with  the regular 3D Leidenfrost  effect is enlightening. In references  \cite{Strier2000},  it has been reported that star-like drop  shapes can be observed when the droplet is trapped and forced to remain at the same place either by the use  of a concave substrate or by the presence of pinning objects such as impurities.  Further  star-like   shapes have also been observed   when  nitrogen drops  are deposited over a viscous liquid at room temperature \cite{Snezhko2008}. In this case, the oscillations seem to be coupled to the  internal flow inside the drop.  More generally the spontaneous oscillation of drop or  liquid puddles 
 seems to be more general and  has also been  observed in other levitated systems without Leidenfrost effect as shown in reference  \cite{Brunet2011}.   In this  previous work,  a liquid droplet or puddle is levitated  over an air cushion which generates  a quasi-laminar flow and star-like oscillations are observed \cite{Brunet2011}.\\
 \begin{figure}[!h]
\includegraphics[width=.4\textwidth]{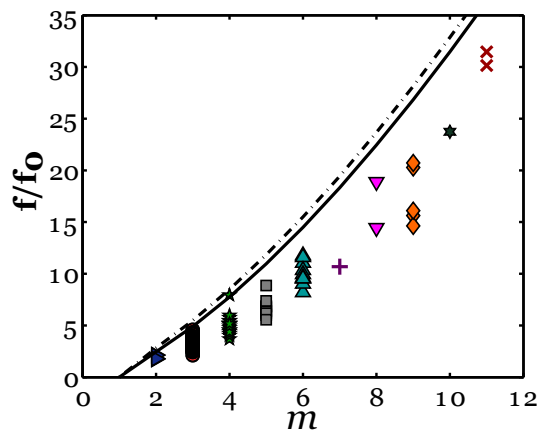}
 \caption{
Dimensionless oscillation frequency $f/f_0$ vs the mode number $m$.  All experiments are plotted ($d=0.3$, 0.5, 1 and 2 mm). For comparison we plot the models of an inviscid 3D spherical drop of the same radius \cite{Lamb1932} ($f^{3D}/f_0 = \sqrt{(m-1) m (m+2)}$, dashed-dotted line) and  for a perfectly homogeneous and discoidal 2D puddle  \cite{Takaki1985} ($f^{2D}/f_0 = \sqrt{m (m^2-1)}$, solid line).}
\label{mode3}
\end{figure}
 It is noteworthy to  mention that in all   theses systems (Leidenfrost  drops 3d and 2d  and levitated air-cushion  drop)  the relevant frequency  of oscillations displays a $R^{-3/2}$ tendency which is a characteristic of Rayleigh capillary wave dispersion relation \cite{Snezhko2008,BouwhuisPreprint}.  
Here again as for our results   (Fig. \ref{mode3})      the measured frequencies seem   to be
slightly overestimated by the  simple capillary wave models \cite{Snezhko2008,BouwhuisPreprint}. 
This fine difference has not been explained yet to our knowledge and  requires as we mention above a mode analysis  with the particular geometry. 


We also  have noticed experimentally  (see movie in the supplementary materials) that contour modes seem to be  associated with  thickness  waves in the confined dimension. This could emphasize the occurrence of non-linear coupling   between   thickness waves and contour waves. Therefore, not only the dispersion relation of the wave shows a small discrepancy as shown in Fig. \ref{mode3}  with the simple models (Rayleigh capilary waves in 2d or 3d) but the origin of the instability leading to star-like droplet  still remains unsolved and is of current investigation.
For example, the origin of the instability of Leidenfrost puddles in non-confined geometry  is still a matter of debate as discussed in reference  \cite{Brunet2011}. It was proposed that oscillations originate from thermal-convection inside the drop \cite{Strier2000}, but observations in athermal systems  show  that  a purely hydrodynamical instability \cite{Brunet2011} is the most probable scenario.
In the case of air-flow levitated droplet, numerical experiments   have been  performed and have revealed a possible scenario for the origin of the instability \cite{BouwhuisPreprint}. In this latter,  the hydrodynamical flow  inside the drop and inside the sustaining  air layer \cite{BouwhuisPreprint} is numerically investigated  and    compared  to experiments.
In both cases, the experiments and  numerical simulations exhibit a critical airflow velocity inside the lubricating layer above which the drop begins to oscillate. A mechanism   has been postulated  where the occurrence of a primary instability (appearance of axi-symetric breathing modes)  triggers  parametrically a second instability and the appearance of the non-axisymetric contours modes observed experimentally   \cite{BouwhuisPreprint}.  For millimeter-sized drops,  it was found  experimentally that this instability appears for airflow velocities higher  than $\sim$ 0.1 m/s . For comparison this value is significantly lower than  the airflow values  inferred from the evaporation  rate in  our Leidenfrost Hele-Shaw  cell.  This could support  the hypothesis of  purely hydrodynamical instability in our experiments as well.

\section{ Conclusion}
We have investigated the behavior of  2d-Leidenfrost droplet in a Hele-Shaw cell. This investigation has revealed a rich behavior of phenomena such has droplet levitation  below and above  a vapor layer, 
and  the appearance of critical transition from discoidal droplet to torus like droplet. This  topological transition occurs due to the thinning of liquid at the center of the droplet.
We have characterized experimentally these phenomena and we have presented a  theoretical 
model  and a numerical study based on the lubrication approximation  which corroborate these experimental
facts.  We are currently investigating the dynamics of the hole expansion.  We also report observations of  large amplitude star-like undulation  consisting of  azimuthal  oscillating capillary waves. The frequency of  oscillations are measured and are found to be close to the frequency of Rayleigh  capillary wave  of droplets. Finally, we  have discussed  possible   mechanisms  which are at the origin of this instability.

\section{acknowledgements}
 We would like to thank Martine Le Berre and Damien Scandola  for  fruitful discussions.

\section{Appendix A }
For the sake of clarity,  we derive here  Eq. (\ref{eqpomeau}).  Let us recall that the temperature in the vapor layer is given by
\begin{equation}
T=T_p \left(1-\frac{z}{h(x,y)}\right)+T_b \frac{z}{h(x,y)} \, .
\end{equation}
The  horizontal velocity  components $u$ and $v$ read using the lubrication approach:

$$u=  \frac{z \left(z -h(x,y)\right)}{2 \eta}  \partial_x   p \,   \, \,   {\rm and}  \,    \,   \, v=  \frac{z \left(z -h(x,y)\right)}{2 \eta}  \partial_y   p \, .$$
Integrating the incompressibility relation: 
$$\partial_x  u +\partial_y  v+ \partial_z  w=0  \, ,$$  between $z=0$  where $w=0$ and $z=h(x,y)$ we obtain using Eq. (\ref{eq1}),

 \begin{equation}\nabla_2 \cdot  \left(\frac{h^3}{12} \nabla_2 p\right) +\frac{ \eta  \kappa \Delta T}{h(x,y)}=0  \, ,
 \label{fulleq}
 \end{equation}

where 
 $\kappa=\lambda/(\rho_v L)$  \, and $\nabla_2= \mathbf{e}_x \partial_x  +\mathbf{e}_y \partial_y$.
 In cylindrical coordinates,  Eq. (\ref{fulleq})
 reads :
 \begin{equation}
 \partial_r \left(\frac{ r h^3 }{12} \partial_r p\right)+\frac{ r \eta \kappa \Delta T}{h(r)}=0 \, .
\end{equation} and is exactly similar to Eq. (\ref{eqpomeau}).

 \end{document}